\documentclass[aps,prb,onecolumn,showpacs,superscriptaddress]{revtex4}
\usepackage[latin1]{inputenc}
\usepackage{amsmath}
\usepackage{amsfonts}
\usepackage{amssymb}
\usepackage{multirow}
\usepackage{graphicx}
\usepackage{epstopdf}
\begin{document}

%\preprint{}
\title{Structure and composition of the superconducting phase in 
alkali iron selenide K$_y$Fe$_{1.6+x}$Se$_2$}
\author{Scott V. Carr}
\affiliation{Department of Physics and Astronomy, Rice University, Houston, Texas 77005, USA}
\author{Despina Louca}
\email{louca@virginia.edu}
\affiliation{Department of Physics, University of Virginia, Charlottesville, Virginia 22904, USA}
\author{Joan Siewenie}
\affiliation{Lujan Neutron Scattering Center, Los Alamos National Laboratory, Los Alamos, New Mexico 87545, USA}
\author{Q. Huang}
\affiliation{NIST Center for Neutron Research, National Institute of Standards and Technology, Gaithersburg, MD 20899, USA}
\author{Aifeng Wang}
\affiliation{Hefei National Laboratory for Physical Science at Microscale and Department of Physics, University of Science and Technology of China,
Hefei, Anhui, 230026, China}
\author{Xianhui Chen}
\affiliation{Hefei National Laboratory for Physical Science at Microscale and Department of Physics, University of Science and Technology of China,
Hefei, Anhui, 230026, China}
\author{Pengcheng Dai}
\email{pdai@rice.edu}
\affiliation{Department of Physics and Astronomy, Rice University, Houston, Texas 77005, USA}

\begin{abstract}
We use neutron diffraction to study the temperature evolution of the  
average structure and local lattice distortions in insulating and superconducting potassium iron selenide K$_y$Fe$_{1.6+x}$Se$_2$.
In the high temperature paramagnetic state, both materials have a single phase with crystal structure similar to that of the BaFe$_2$As$_2$ family of  
iron pnictides. While the insulating K$_y$Fe$_{1.6+x}$Se$_2$ forms  
a $\sqrt{5}\times\sqrt{5}$ iron vacancy ordered block antiferromagnetic (AF) structure 
at low-temperature, the superconducting compounds spontaneously phase separate into 
an insulating part with $\sqrt{5}\times\sqrt{5}$ iron vacancy order and a superconducting phase with chemical composition of 
K$_z$Fe$_{2}$Se$_2$ and BaFe$_2$As$_2$ structure. 
Therefore, superconductivity in alkaline iron selenides arises from alkali deficient K$_z$Fe$_{2}$Se$_2$ in the matrix of the insulating block AF phase.
\end{abstract}


\maketitle

\section{Introduction}
Of all the iron-based superconductors \cite{johnston,stewart,dai}, alkali iron selenides $A_y$Fe$_{1.6+x}$Se$_2$  ($A = $ K, Rb, Cs, Tl) \cite{jgguo,krzton,mhfang,afwang,haggstrom,dagotto}
are unique in that superconductivity in this class of materials always coexists with a static long-range antiferromagnetic (AF) order with a large moment and high 
N$\rm \acute{e}$el temperature \cite{wbao1,pomjakushin1,fye,mwang11,despina}.  This is in contrast to iron pnictide superconductors \cite{johnston,stewart,dai} where optimal superconductivity arises from the 
 suppression of the static AF order in their nonsuperconducting 
 parent compounds \cite{kamihara,cruz}.  An attempt to understand the 
 coexisting static AF order and superconductivity in alkali iron selenides \cite{wbao1,pomjakushin1,fye,mwang11,despina} has produced two proposed scenarios. In the   
first, superconductivity is believed to coexist microscopically with the static AF order \cite{wbao1,despina}.  However, the AF order in $A_y$Fe$_{1.6+x}$Se$_2$ forms a 
$\sqrt{5}\times\sqrt{5}$ block AF structure with an ordered moment of $\sim$3.3 $\mu_B$ per Fe as shown in Fig. 1(a) \cite{wbao1,pomjakushin1,fye,mwang11,despina}, making it 
unclear how superconductivity can survive such a large magnetic field background arising from the ordered moments \cite{mazin2011}. Alternatively, 
superconductivity in $A_y$Fe$_{1.6+x}$Se$_2$ may arise from a chemically separated superconducting phase in the matrix of the insulating  
block AF phase. Although transmission electron microscopy (TEM) \cite{zwwang,speller12,xxding,zwwang13}, X-ray/neutron diffraction \cite{mwang11,ricc2,jzhaoprl,shoemaker}, scanning
tunneling microscopy (STM) \cite{wli11}, M$\rm \ddot{o}$ssbauer spectroscopy \cite{Ksenofontov}, 
muon spin relaxation \cite{Shermadini12}, apertureless scattering-type scanning near-field optical microscopy \cite{Charnukha},
and nuclear magnetic resonance \cite{Texier} experiments have provided ample evidence for
phase separation, where superconductivity comprises about 10-20\% of the volume of $A_y$Fe$_{1.6+x}$Se$_2$, there is currently no consensus on the chemical 
composition or crystal structure for the superconducting $A_y$Fe$_{1.6+x}$Se$_2$. For example, while some TEM \cite{zwwang13},  X-ray scattering \cite{shoemaker}, and STM \cite{wli11}  measurements suggest
that the superconducting phase of $A_y$Fe$_{1.6+x}$Se$_2$ is $A_z$Fe$_2$Se$_2$ with the BaFe$_2$As$_2$ 
iron pnictide crystal structure \cite{johnston,stewart,dai}, other TEM and STM measurements propose that
the superconducting phase consists of a single
Fe vacancy for every eight Fe-sites arranged in a $\sqrt{8}\times\sqrt{10}$ parallelogram structure [Fig. 1(c)] \cite{xxding}.  In addition, single crystal neuron diffraction experiments 
indicate that superconductivity in $A_y$Fe$_{1.6+x}$Se$_2$ may arise from
a semiconducting AF phase with rhombus iron vacancy order [Fig. 1(b)] \cite{jzhaoprl} 
instead of the well-known the insulating $\sqrt{5}\times\sqrt{5}$ block AF phase \cite{wbao1,pomjakushin1,fye,mwang11}.

\begin{figure}[h]
\includegraphics[scale=.48]{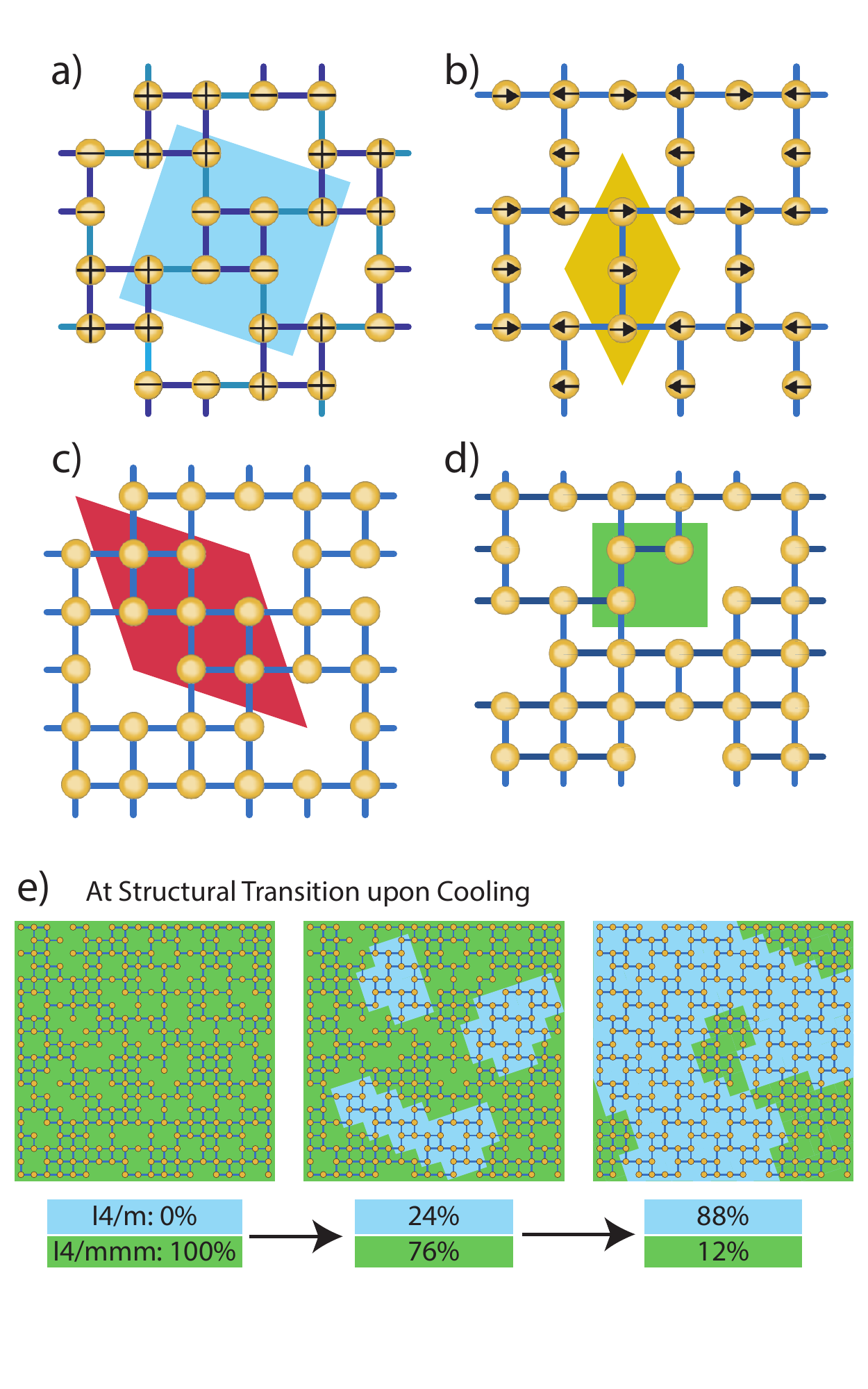}
\caption{(color online) 
In plane structures proposed for various $A_y$Fe$_{1.6+x}$Se$_2$ phases: a) Main vacancy ordered phase with $\sqrt{5}\times\sqrt{5}$ block AF order 
\cite{wbao1,pomjakushin1,fye,mwang11}. The $+$ and $-$ signs indicate Fe spin directions relative to the FeSe plane.
b) Proposed phase for semiconductor, where spin directions are marked as arrows \cite{jzhaoprl}. c) Suggested superconducting phase with 
$\sqrt{8}\times\sqrt{10}$ iron vancancy order \cite{xxding}. 
d) Iron-disordered, partially occupied orthorhombic phase with BaFe$_2$As$_2$ iron pnictide structure. e-g) 
Schematic for phase separation through spontaneous nucleation in the temperature region of iron mobility.  A disordered $I4/mmm$ phase above 
$T_s$. Below $T_N $, the $I4/m^\prime$ symmetry insulating phase forms at random sites and spreads 
enriching the iron in remaining disordered phase until either full iron occupation of second phase (slow cooling) or temperature drops below zone of iron mobility (quenching).
}
\end{figure}

Given the numerous proposed crystal structures for the superconducting $A_y$Fe$_{1.6+x}$Se$_2$, a
determination of the true chemical composition and relationship with the AF insulating phase is essential to
understand the bulk electronic properties \cite{dagotto}. In this article, we present systematic
neutron powder diffraction measurements on superconducting and insulating K$_y$Fe$_{1.6+x}$Se$_2$.  By carefully comparing 
Rietveld refinements of neutron diffraction spectra of the superconducting and insulating K$_y$Fe$_{1.6+x}$Se$_2$ in the high temperature 
paramagnetic phase, we find that both materials are phase pure with the 
iron pnictide crystal structure (space group $I4/mmm$) \cite{johnston} and slightly more iron in the superconducting sample. On cooling to the low-temperature ground state, while  
the insulating K$_y$Fe$_{1.6+x}$Se$_2$ remains a single phase, now of a $\sqrt{5}\times\sqrt{5}$ iron vacancy structure with block AF order \cite{wbao1,pomjakushin1,fye,mwang11},
the superconducting K$_y$Fe$_{1.6+x}$Se$_2$ becomes phase separated into the stoichiometric $\sqrt{5}\times\sqrt{5}$ iron vacancy ordered K$_{0.8}$Fe$_{1.6}$Se$_2$ and 
potassium deficient K$_z$Fe$_{2}$Se$_2$ 
phase with the iron pnictide crystal structure \cite{johnston,stewart,dai}.  The superconducting phase arises in the region
of the excess iron between the low temperature
$\sqrt{5}\times\sqrt{5}$ iron vacancy ordered phase, and its volume fraction can be controlled through temperature cycling and quenching processes.  Rietveld and pair distribution function (PDF) 
analysis of the data indicates that the superconducting phase cannot be the stoichiometric $A$Fe$_2$Se$_2$ \cite{zwwang13,wli11}, 
$\sqrt{8}\times\sqrt{10}$ \cite{xxding} or  rhombus \cite{jzhaoprl} iron vacancy ordered phase.  Therefore, the 
superconducting phase in $A_y$Fe$_{1.6+x}$Se$_2$ arises from a spontaneous chemical phase separation 
from the insulating $\sqrt{5}\times\sqrt{5}$ iron vacancy ordered phase 
due to excess iron, fundamentally different from superconductivity in iron pnictides \cite{zwwang13,shoemaker}.

\begin{figure}[h]
\includegraphics[scale=.45]{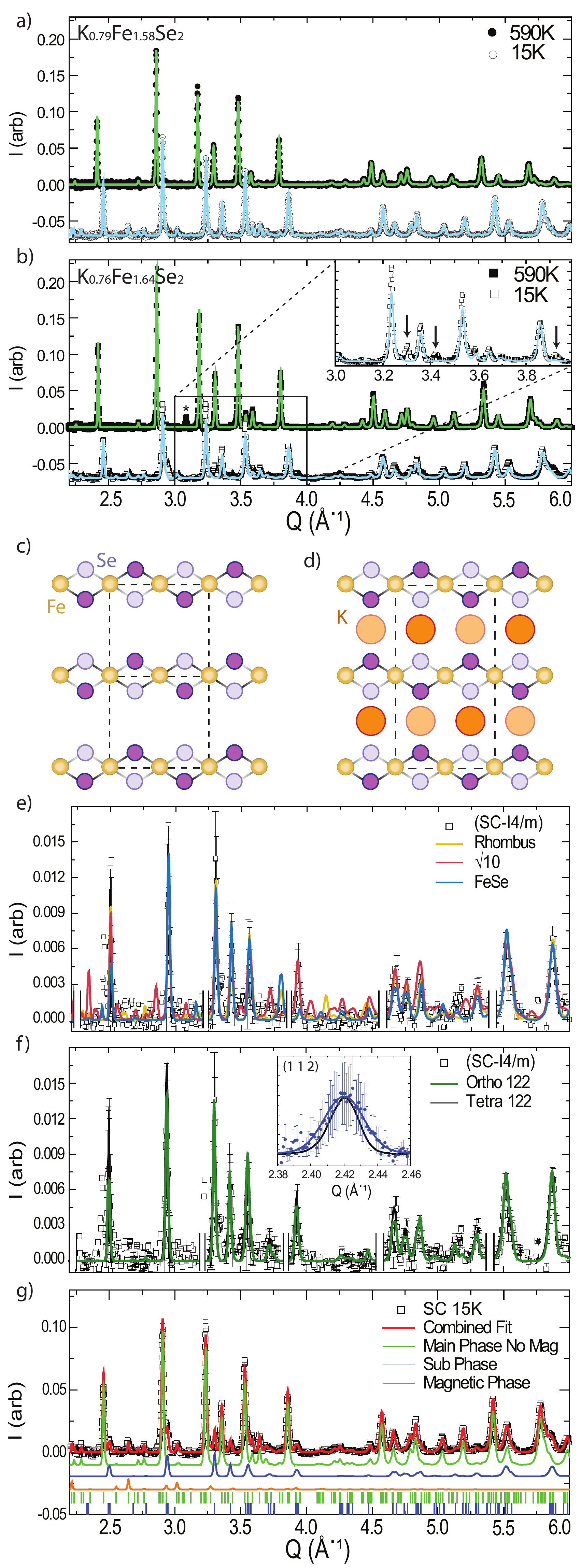}
\caption{(color online) Region of powder diffraction spectra with corresponding single phase fits for the a) insulating and b) superconducting sample, each fit with an $I4/mmm$ symmetry phase at 
590 K and $I4/m^\prime$ symmetry phase at 15 K.  (b inset) The unfit peaks in the superconducting sample with only the $I4/m^\prime$ 
symmetry are marked with arrows.  There is an unidentified impurity peak near 3.2 \AA$^{-1}$\ marked as $\ast$ in the superconducting sample. The peak has no temperature
dependence. Side cut of pure c)FeSe and d)KFe$_2$Se$_2$ crystal structures. Darker atoms are placed in front of iron while lighter ones behind. 
Note inversion of middle layer necessary to form the BaFe$_2$As$_2$ symmetry phase from FeSe. e) Best fits using different proposed second phases.  
FeSe can be interpreted as a K-intercalated sample with no K ordering. f) Best fit using orthorhombic $Fmmm$ symmetry structure 
akin to that of BaFe$_2$As$_2$. The inset shows clear broadening of the $(1 1 2)$ peak at base temperature (blue dots) compared to above $T_s$ (line).
Such broaden can be induced by a small orthorhombic lattice distortion. g) Comparison of the fits using pure $\sqrt{5}\times\sqrt{5}$ iron vacancy order
without AF order (green), the effect of AF order (orange), the superconducting phase (blue), and the combination of all phases (red).  At 15 K, the lattice
parameters of the block AF phase for the superconducting and insulating samples are  $a=8.68535(6)$ \AA, $c=14.0054(2)$ \AA\ and 
$a=8.68403(5)$ \AA, $c=14.0084(2)$ \AA, respectively.  The lattice parameters for the K$_z$Fe$_{2}$Se$_2$ phase are 
$a=5.3966(3)$, $b=5.3653(3)$, and $c=14.043(1)$ \AA.
}
\end{figure}

\begin{figure}[h]
\includegraphics[scale=.7]{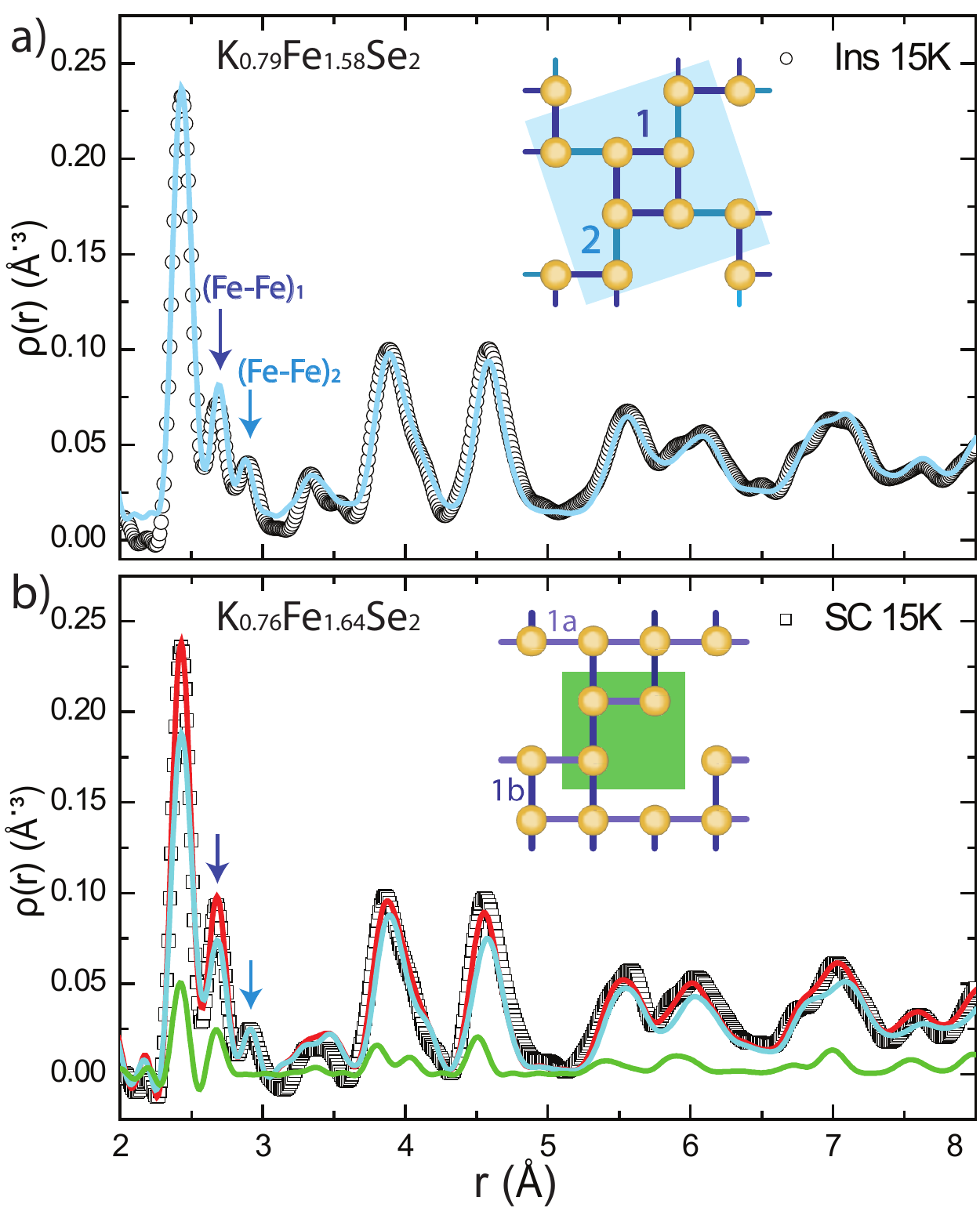}
\caption{(color online) Pair Distribution Function at base temperature of the insulating and superconducting samples. a) The insulator at base temperature is fit using only the $I4/m^\prime$ symmetry phase (inset, blue line).  The split in the Fe-Fe peak results from a contraction of [Fe-Fe]$_1$ and rotation of inner iron square. b) The superconductor requires an addition phase, here fit with $Fmmm$ symmetry (inset, green line) and disordered Fe vacancies.  The Fe-Fe bond distance in the additional phase matches the [Fe-Fe]$_1$ distance, resulting in an increase of intensity relative to [Fe-Fe]$_2$.
}
\end{figure}

\section{Experimental Results and Discussions}
We carried out neutron scattering experiments using the neutron time of flight powder diffractometer NPDF at the Los Alamos National laboratory.
This spectrometer has the advantage of a large $Q$-range, allowing a determination of the 
average as well as local structures of the system through Rietveld \cite{mwang11} and PDF analysis of the diffraction spectra \cite{despina}, respectively.
We grew  several large single batches of K$_y$Fe$_{1.6+x}$Se$_2$
via the self-flux method with nominal dopings of $y=0.8$, $x=0$ and $y=0.8$, $x=0.4$ for the insulating and superconducting samples, respectively. 
Since transport measurements are sensitive to paramagnetic tetragonal to iron vacancy ordered structural phase transition ($T_s$), AF ordering ($T_N$), and superconductivity $T_c$ \cite{dagotto}, 
samples from each batch were characterized by in-plane resistivity and magnetization to determine $T_s$, $T_N$, and $T_c$ to be approximately 540 K, 500 K and 32 K in the superconducting sample and 520 K and 500 K in the insulator.  From previous neutron diffraction work \cite{wbao1,pomjakushin1,fye,mwang11}, we know that structural and magnetic phase transitions in 
K$_y$Fe$_{1.6+x}$Se$_2$ occur above the room temperature.  Choosing only crystals from a single batch to minimize stoichiometry differences, we 
ground several grams of each composition into a fine powder and divided it into two sets.  One set of powders was measured only at low temperatures while the other was measured to temperatures above 
$T_s$ to observe any hysteretical effects. For the neutron diffraction experiments, 
each sample was effectively measured along a temperature loop beginning at the base temperature of 15 K and measured at intervals to above $T_s$ before returning to room temperature. 
This produced two sets of data, one for superconducting and one for insulating K$_y$Fe$_{1.6+x}$Se$_2$, at identical temperatures to enable direct comparison. 
Following the completion of the neutron scattering 
measurements, exact crystal stoichiometry from small portions of both 
high and low temperature powder was measured via Inductively Coupled Plasma (ICP) analysis. 

\begin{figure}[t]
\includegraphics[scale=.4]{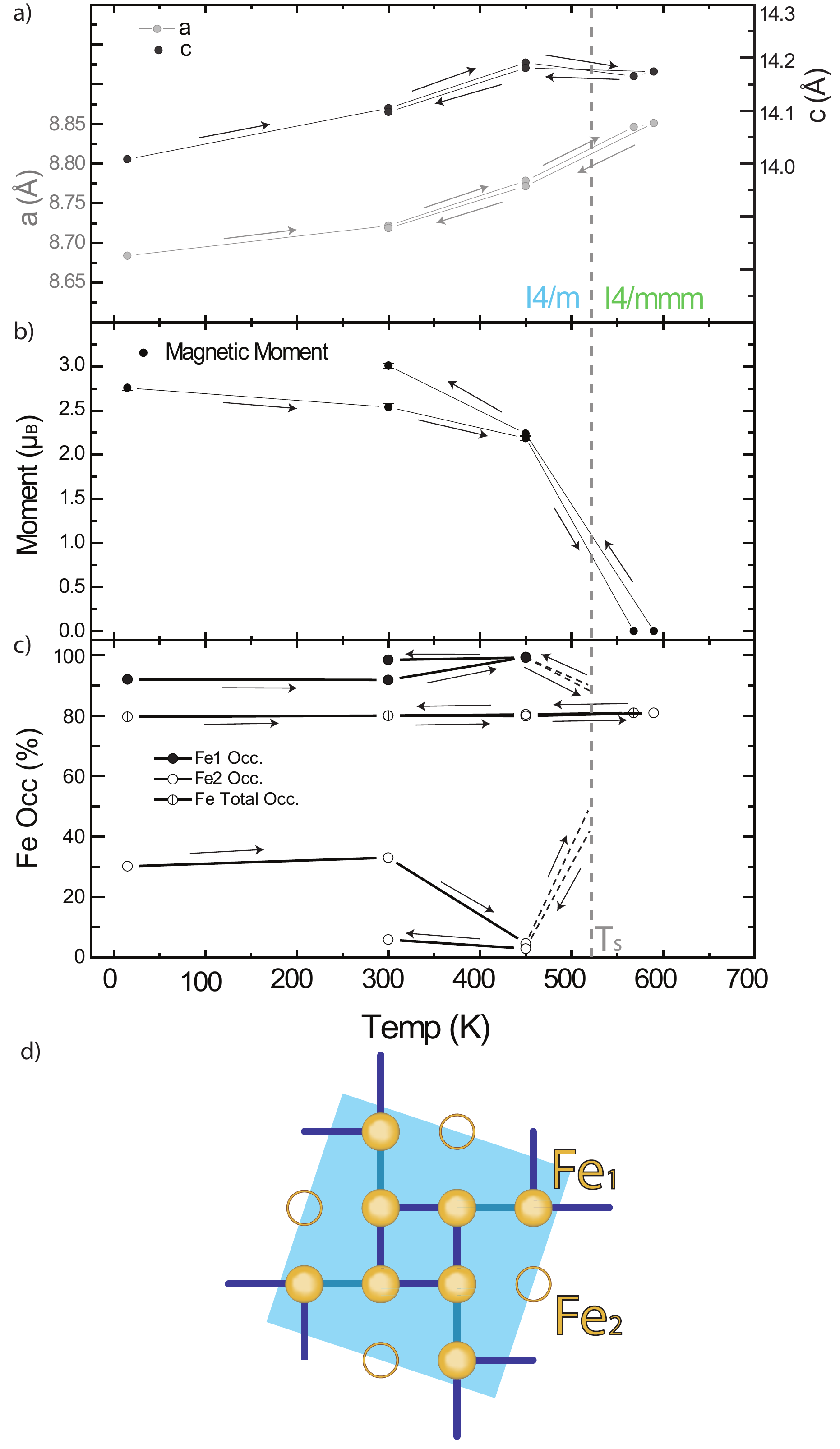}
\caption{(color online) a) Refined lattice parameters for the insulating sample.  All results are reported using the $I4/m^\prime$ symmetry unit cell. b) Refined magnetic 
moment shows hysteresis and strengthens upon annealing. c) Refined iron concentration for iron occupation by site also shows enhanced vacancy ordering below 450 K upon warming 
from a quenched state. d) Iron site labels in the $I4/m^\prime$ symmetry phase. All arrows indicate direction in thermal cycle.
}
\end{figure}

We carried out Rietveld analysis to determine the average crystal 
structure and chemical compositions of the superconducting and insulating 
K$_y$Fe$_{1.6+x}$Se$_2$.  Figure 2(a) shows the intensity versus wave vector spectra 
and fits from the Rietveld analysis for the insulating K$_y$Fe$_{1.6+x}$Se$_2$.
In the paramagnetic tetragonal state (top panel) at $T=590$ K, the spectra can be well fit by 
the $I4/mmm$ symmetry tetragonal unit cell (green solid line) suitable 
for the paramagnetic tetragonal state of BaFe$_2$As$_2$ \cite{johnston}.
On cooling to $T=15$ K in the block AF ordered state, we find that the spectra can still be well described by a 
single phase of K$_{0.78}$Fe$_{1.58}$Se$_2$, but with the space group  $I4/m^\prime$ and
$\sqrt{5}\times\sqrt{5}$ iron vacancy order [Fig. 1(a)]
 \cite{wbao1,pomjakushin1,fye,mwang11}. Therefore, there is no phase separation in the insulating 
K$_{0.78}$Fe$_{1.58}$Se$_2$.  

Figure 2(b) shows neutron diffraction spectra for the superconducting K$_y$Fe$_{1.6+x}$Se$_2$.
In the paramagnetic tetragonal phase ($T=590$ K), we find that a single phase tetragonal unit cell with 
the $I4/mmm$ symmetry can fit the data well [top panel in Fig. 2(b)].  
The refined stoichiometry indicates that the 
superconducting K$_y$Fe$_{1.6+x}$Se$_2$ has more iron ($x=0.029$) than that of the insulator ($x=0.009$).
On cooling to $T=15$ K, we find  
that the $I4/m^\prime$ space group suitable for the insulating K$_{0.78}$Fe$_{1.58}$Se$_2$ in Fig. 2(a) 
can still fit majority of the Bragg peaks very well.  However, 
there are additional Bragg peaks occurring at positions not allowed by the $I4/m^\prime$ symmetry [bottom panel and 
arrows in the inset of Fig. 2(b)].

\begin{table}[h]
\centering
\begin{tabular}{|c|c|c|c|c|c|c|}
\hline
T =15K & Atom & $x$ & $y$ & $z$ & $U_{iso}$ (10$^{-2})$ & Occ\\
\hline
\multirow{6}{*}{Phase 1 (86.39\%)} & K$_1$ & 0 & 0 & 0 & 1.58(9) & 0.797(8)\\
& K$_2$ & 0.3944 & 0.2141 & 0 & 1.58(9) & 0.797(8)\\
& Fe$_1$ & 0 & 0.5 & 0.25 & 0.42(2) & 0.32(1) \\
& Fe$_2$ & 0.1989 & 0.0915 & 0.2532 & 0.42(2) & 0.913(4) \\
& Se$_1$ & 0.5 & 0.5 & 0.1384 & 0.51(2) & 1 \\
& Se$_2$ & 0.1110 & 0.3103 & 0.1443 & 0.51(2)& 1\\
\hline
\multirow{3}{*}{Phase 2 (13.61\%)} & K & 0 & 0 & 0 & 2.65(9) & 0.53(3)\\
& Fe & 0.25 & 0.25 & 0.25 & 0.51(1) & 1.00(2)\\
& Se & 0 & 0 & 0.3553 & 0.32(2)& 1\\
\hline
\end{tabular}
\caption{
Phase 1 is the block AF phase and Phase 2 is the iron pnictide phase with K deficiency.   
$x$, $y$, and $z$ are positions of atoms in the unit cell, and $U_{iso}$ is the isotropic
Debye Waller factor. Occ is the occupancy of the atomic site for different atoms.
}
\end{table}

Beginning with the assumption that the new phase is associated with superconductivity, there are many possible
crystalline structures. These include the rhombus iron vacancy order [Fig. 1(b)] \cite{jzhaoprl}, 
$\sqrt{8}\times\sqrt{10}$ iron vacancy ordered phase [Fig. 1(c)] \cite{xxding}, 
iron disordered $I4/mmm$ tetragonal symmetry [Fig. 1(d)], expanded structure of FeSe [Fig. 2(c)] \cite{wu,tmmcqueen,tkchen}, and finally the BaFe$_2$As$_2$ type 
iron pnictide structure [Fig. 2(d)] \cite{johnston,stewart,dai}. To determine the space group and chemical composition of the new phase in the low temperature state, we 
plot in Fig. 2(e) the difference plot between the superconducting and insulating $K_y$Fe$_{1.6+x}$Se$_2$, 
omitting regions of peak misalignment due to slightly different lattice parameters (See Fig. 2 caption). We can then use different proposed models for the superconducting 
$K_y$Fe$_{1.6+x}$Se$_2$ to fit the observed spectra.
The yellow, red, and blue solid lines in Fig. 2(e) show Rietveld fits of the data using the 
rhombus \cite{jzhaoprl}, $\sqrt{8}\times\sqrt{10}$ \cite{xxding} iron vacancy ordered, and 
pure FeSe \cite{wu,tmmcqueen,tkchen} structures, respectively.  None of them can match allowed peak position and intensity properly.
Figure 2(f) shows Rietveld analysis using the BaFe$_2$As$_2$ structure.  Although the tetragonal
$I4/mmm$ symmetry suitable for the high temperature paramagnetic state 
fits the data fairly well (black line), we find that a slight orthohombic distortion (space group $Fmmm$) fits the data better as shown
in solid green line of Fig. 2(f).  A direct comparison of   
the low (blue circles) and high temperature (solid line) data in the inset of Fig. 2(f) displays slight broadening of the low-temperature data, confirming the low-temperature orthorhombic symmetry.
The best overall fit yielded 13.6\% of the crystal volume in the second phase with a refined stoichiometry of K$_{0.53}$Fe$_{2.002}$Se$_2$, and the remainder in 
the block AF phase.  The combined fit and Bragg peak positions for different phases are shown in Fig. 2(g) and positions of all atoms in the unit cell are summarized in Table I.
Using independently fit stoichiometry and phase fraction, we can compare a calculated overall crystal stoichiometry with the 
measured ICP values.
Doing so at base temperature results in an overall crystal stoichiometry of K$_{0.75\pm0.01}$Fe$_{1.64\pm0.02}$Se$_2$ providing 
excellent agreement with ICP measurements of 
K$_{0.76}$Fe$_{1.64}$Se$_2$.

\begin{figure}[t]
\includegraphics[scale=.4]{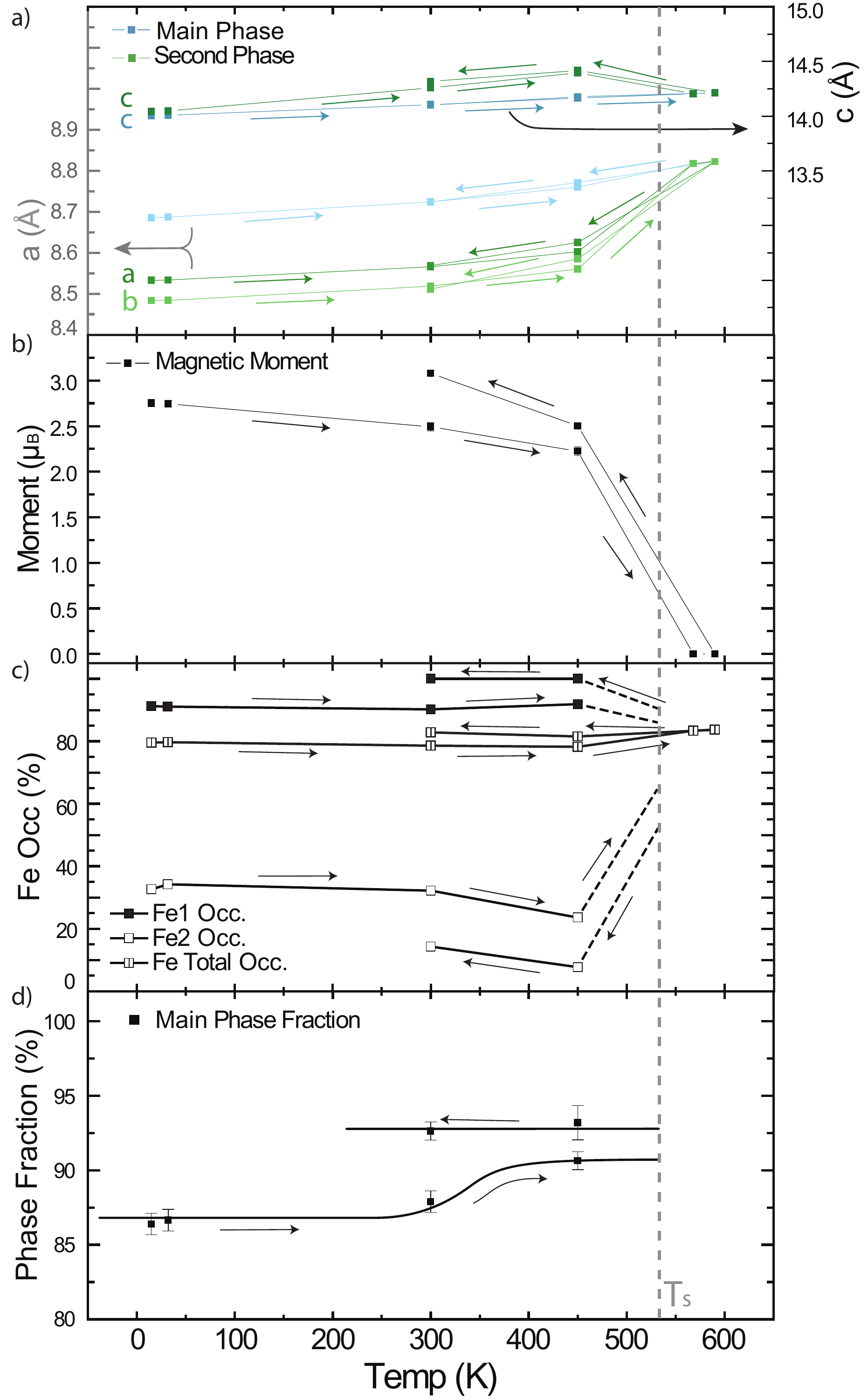}
\caption{(color online) a) Refined lattice parameters for both insulating (blue) and superconducting (green) phases.  Second phase features notable compression (~2\% at 15 K), 
orthorhombicity (0.29\%), and hysteresis along the $c$-axis. b) Refined magnetic moment in insulating phase showing same behavior as pure insulator. c) Refined iron occupation in insulating phase by site.  Notable refinement of vacancy order upon temperature cycling. d) Refined main phase fraction upon temperature cycling.  A clear increase in main phase fraction upon warming to 450 K in a quenched sample. Solid lines are guides to the eye.
}
\end{figure}

In addition to carrying out Rietveld analysis, we can also compare the low-temeprature local structural distortion of the superconducting and insulating $K_y$Fe$_{1.6+x}$Se$_2$ using 
the PDF calculation.  At base temperature, the $\sqrt{5}\times\sqrt{5}$ iron vacancy order is already established and 
the square of iron in the center of the unit cell [see inset in Figure 3(a)] twists and contracts, leading to a slightly different 
intra- (light) and inter-square (dark) Fe-Fe bond distance \cite{wbao1,pomjakushin1,fye,mwang11}. Indeed, this is observed as two marked peaks
near 2.7 angstroms in the PDF analysis of the insulating K$_y$Fe$_{1.6+x}$Se$_2$ [Fig. 3(a)], where the integrated peak intensity indicates the population 
density of atomic pairs in the sample.  Figure 3(b) shows the same PDF analysis for the superconducting $K_y$Fe$_{1.6+x}$Se$_2$. Although we still see the same splitting of the
two Fe-Fe bond distances, the intensity ratio of the split peaks differs between compositions. 
For a pure $\sqrt{5}\times\sqrt{5}$ iron vacancy phase with $I4/m^\prime$ symmetry, one would expect integrated intensity of these peaks to be in a 2:1 ratio reflecting 
the number of such atomic pairs in the unit cell [Fig. 3(a)].  While a quantitative comparison is difficult, a qualitative comparison to the superconducting and insulating samples is still instructive. 
In Figure 3(b) for the superconducting sample, we see that  
the intensity of [Fe-Fe]$_1$ is substantially larger compared to the same peak in the insulator despite comparable intensity of the [Fe-Fe]$_2$ peak. 
This discrepancy can only be understood assuming the presence of an additional 
phase in the superconducting sample with one Fe-Fe bond distance, consistent with Rietveld analysis shown in Fig. 2.

If we assume that the superconducting K$_y$Fe$_{1.6+x}$Se$_2$ is phase pure in the high temperature paramagnetic state,
it would be interesting to understand how it becomes phase separated at low temperatures.  In a recent TEM paper \cite{zwwang13}, it was argued that the superconducting phase
in $K_y$Fe$_{1.6+x}$Se$_2$ has a chemical composition of K$_{0.5}$Fe$_2$Se$_2$ and originates from spinodal phase separation around $T\approx 540$ K to form a Archimedean solid-like framework 
embedded in the insulating K$_{0.8}$Fe$_{1.6}$Se$_2$ matrix.  These results are consistent with previous X-ray diffraction findings which suggest phase separation \cite{shoemaker}.
While this scenario is not in conflict with our data, we present another possible picture in Fig. 1(e)-1(g) showing the temperature 
evolution of the crystal structures for superconducting K$_y$Fe$_{1.6+x}$Se$_2$. At high temperature, the entire sample is in the high symmetry $I4/mmm$ phase. 
At this temperature, the iron atoms are extremely mobile resulting in a homogeneous iron deficient iron pnictide structure [Fig. 1(e)]. 
When cooled below $T_s$, a $\sqrt{5}\times\sqrt{5}$ iron vacancy ordered phase with chemical composition 
K$_{y}$Fe$_{1.6}$Se$_2$ forms due to its lower energy \cite{txiang,xwyan,wgyin}. 
The excess iron in the high temperature paramagnetic tetragonal phase are segregated from the 
iron vacancy ordered main phase, resulting in
K$_{y}$Fe$_2$Se$_2$, which retains the iron pnictide structure [Fig. 1(f) and 1(g)]. In this picture, the rate of cooling and average Fe concentration will determine the final phase fractions as well as second phase stoichiometry. This picture is consistent with the hysteresis effects in superconducting volume fraction seen in
the scanning electron microscopy measurements \cite{yliu12}.

To quantitatively determine the effect of annealing on the insulating and superconducting of K$_y$Fe$_{1.6+x}$Se$_2$, 
we carried out careful Rietveld analysis of the temperature evolution of the neutron diffraction spectra.
Figure 4 summarizes the outcome for the insulating K$_{0.78}$Fe$_{1.58}$Se$_2$.  In an ideal $\sqrt{5}\times\sqrt{5}$ block AF structure, 
there should be full occupancy at iron site 1 (Fe$_1$) and full vacancy at iron site 2 (Fe$_2$) as shown in Fig. 4(d) \cite{wbao1,pomjakushin1,fye,mwang11}.
Figure 4(a) shows no hysteresis in 
temperature dependence of the lattice parameters across the block AF ($I4/m^\prime$) to paramagnetic tetragonal ($I4/mmm$) structural phase 
transition. Figure 4(b) shows temperature dependence of the ordered magnetic moment, which clearly increases at 300 K 
after warming up above $T_s$. From Fig. 4(c), we see that the significant iron occupation ($\sim$30\%) in the as-grown K$_{0.78}$Fe$_{1.58}$Se$_2$ almost vanishes
upon warming up to 450 K, suggesting a wide temperature region of iron mobility, enabling an ideal $\sqrt{5}\times\sqrt{5}$ block AF structure that increases the Fe ordered moment.

Figure 5 summarizes similar Rietveld analysis for the superconducting K$_y$Fe$_{1.6+x}$Se$_2$. 
The  lattice parameters of the block AF and iron pnictide phases merge together above $T_s$, consistent with a pure paramagnetic tetragonal phase [Fig. 5(a)].
Figure 5(b) shows temperature dependence of the ordered moments for the block AF structure, which again reveals a clear increase at 300 K 
after warming up to above $T_c$ similar to the insulating K$_{0.78}$Fe$_{1.58}$Se$_2$ [Fig. 4(b)].  Temperature dependence of the Fe$_1$ and Fe$_2$
occupancies for the block AF phase is shown in Fig. 5(c), behaving again similarly as that of the insulating K$_{0.78}$Fe$_{1.58}$Se$_2$ [Fig. 4(c)].
Figure 5(d) plots the temperature dependence of the volume fraction for the block AF phase.  We see that the annealing process slightly increases the
insulating volume fraction at the expense of the superconducting phase 
in K$_y$Fe$_{1.6+x}$Se$_2$.  Comparison of the lattice parameters of the superconducting K$_y$Fe$_{1.6+x}$Se$_2$ with those of FeSe under pressure (Table II) 
reveals that the superconducting K$_y$Fe$_{1.6+x}$Se$_2$ has lattice parameters 
similar to those of compressed FeSe \cite{Medvedev}. Furthermore, the in-plane compression in the second phase and its corresponding $T_c$ fits well with the trend in FeSe.  This may explain the enhanced $T_c$ of K$_y$Fe$_{1.6+x}$Se$_2$ compared with FeSe and suggests that superconductivity in these systems is related.

\begin{table}[h]
\centering
\begin{tabular}{ | c | c | c | c | c | }
\hline
\multicolumn{5}{|c|}{In plane lattice parameter compression} \\
\hline
Compound & Pressure/Temperature & $a$ (ortho) & $\Delta a$\% & $T_c$ (K) \\
\hline
\multirow{5}{*}{FeSe} & 0.25 GPa & 5.2952 & 0.0\% &  10 \\
& 4 GPa & 5.1749 & 2.3\% &  28 \\
& 6 GPa & 5.1010 & 3.5\% &  37 \\
& 9 GPa & 5.0835 & 4.0\% &  23 \\
& 12 GPa & 5.0470 & 4.7\% &  11 \\
\hline
\multirow{2}{*}{K$_z$Fe$_{2}$Se$_2$} & 590 K & 5.580 & 0.2\% & - \\
& 15 K & 5.381 & 3.9\% & 32 \\
\hline
\end{tabular}
\caption{Comparison of the in-plane lattice parameters of FeSe under pressure \cite{Medvedev} and 
that of the superconducting K$_z$Fe$_{2}$Se$_2$ relative to room temperature BaFe$_2$As$_2$ \cite{Huang}.
}
\end{table}

\section{Summary and conclusions}

In summary, we have used neutron scattering to systematically study insulating and superconducting K$_y$Fe$_{1.6+x}$Se$_2$.  By carefully comparing the observed diffraction spectra at different temperatures, 
we conclude that both the insulating and superconducting samples are phase pure in the paramagnetic state with the tetragonal iron pnictide structure.  
While the insulating sample with the chemical composition of K$_{0.78}$Fe$_{1.58}$Se$_2$ is still phase pure in the low temperature state and forms
a $\sqrt{5}\times\sqrt{5}$ block AF structure, the superconducting  K$_{0.76}$Fe$_{1.64}$Se$_2$ phase separates into the insulating K$_{0.798}$Fe$_{1.59}$Se$_2$ with block AF structure 
and superconducting  K$_{0.53}$Fe$_{2}$Se$_2$ with a weakly orthorhombic iron pnictide structure. 
The crystal structure of the superconducting phase in K$_y$Fe$_{1.6+x}$Se$_2$ is consistent with 
the crystal structure of alkali-ammonia intercalated superconducting FeSe compounds if one neglects the ammonia-molecular constituent \cite{burrard,tpying}.
This suggests that the superconducting phase in K$_y$Fe$_{1.6+x}$Se$_2$ and ammonia-intercalated FeSe compounds has the same crystalline structure. 
Our temperature dependent measurements of crystal structural and volume hysteresis suggest that 
phase separation is an intrinsic phenomenon 
in the iron rich K$_y$Fe$_{1.6+x}$Se$_2$.  Therefore, superconductivity appears in the K$_{0.76}$Fe$_{1.64}$Se$_2$ phase with iron pnictide structure, and does not occur in the 
$\sqrt{8}\times\sqrt{10}$ \cite{xxding} or  rhombus \cite{jzhaoprl} iron vacancy ordered phase.

\section{Acknowledgments} 
The work at Rice University is supported by the US NSF DMR-1308603 and OISE-0968226.
The work at the University of Virginia is 
supported by the U. S. DOE, Office of Basic Energy Sciences, under contract number DE-FG02-01ER45927.
This work has benefited from the use of NPDF at the Lujan Center at Los Alamos Neutron Science Center, funded by DOE Office of 
Basic Energy Sciences.  Los Alamos National Laboratory is operated by Los Alamos National Security LLC under DOE 
Contract DE-AC52-06NA25396. The upgrade of NPDF has been funded by NSF through grant DMR 00-76488. Work at USTC is supported 
by MOST and NSFC.

%\bibliography{NoEndingPoint}

\end{document}